\documentclass[11pt, letterpaper]{article}

\usepackage[margin=1in]{geometry}
\usepackage[utf8]{inputenc}
\usepackage[T1]{fontenc}
\usepackage{url}
\usepackage{booktabs}
\usepackage{amsfonts}
\usepackage{amsmath}
\usepackage{graphicx}
\usepackage{microtype}
\usepackage{float}
\usepackage{listings}
\usepackage{siunitx}
\usepackage[numbers,sort&compress]{natbib}
\usepackage{hyperref}
\title{Sequence-Aware Split Heuristic to Mitigate SM Underutilization in FlashAttention-3 Low-Head-Count Decoding}
\author{
    Martí Llopart Font$^{1}$, Javier Hernando$^{1,2}$, Cristina España-Bonet$^{1,3}$\\[0.5em]
    $^{1}$Barcelona Supercomputing Center (BSC-CNS)\\
    $^{2}$Universitat Politècnica de Catalunya\\
    $^{3}$DFKI GmbH, Saarland Informatics Campus\\[0.5em]
    \texttt{mllopart@bsc.es, javier.hernando@bsc.es, cristina.espana@bsc.es}
}
\date{}

\begin{document}
\maketitle

\begin{abstract}
The standard FlashAttention-3 heuristic exhibits a GPU occupancy bottleneck in low-head-count decoding configurations because it disables sequence splitting based on sequence length alone, underutilizing the Streaming Multiprocessors of Hopper GPUs. Our proposed sequence-aware split policy mitigates this by allowing sequence-level parallelism in low-head-count regimes, improving hardware utilization to deliver roughly a 21 to 24\% improvement in decoder kernel efficiency on metadata-enabled inference paths, with no observed regressions.
\end{abstract}

\section{Introduction}
\label{sec:intro}
Modern Large Language Models (LLMs) increasingly use low-key/value-head attention variants during autoregressive decoding. In Multi-Query Attention (MQA), all query heads share a single key/value head \cite{shazeer2019mqa}; in Grouped-Query Attention (GQA), groups of query heads share each key/value head \cite{ainslie2023gqa}. These designs reduce the size of the key-value (KV) cache, i.e., the stored keys and values from previous tokens, but also reduce the parallel workload available per decode step when the number of KV heads $H_{KV}$ is small. This paper focuses on that low-head-count regime, including pure MQA and GQA configurations with few KV heads, particularly when context lengths ($L_K$) are small ($\le 512$).

On highly parallel architectures like the NVIDIA H100 (Hopper), which features 132 Streaming Multiprocessors (SMs), this reduction in concurrent tiles exposes a performance bottleneck. The default dispatch heuristic in FlashAttention-3 \cite{shah2024flashattention3} aborts sequence splitting in these low-tile regimes. As a result, the kernel launches as few as 8 Thread Blocks (CTAs), leaving over 90\% of the GPU SMs idle.

In this paper, we introduce a sequence-aware split heuristic that identifies low-occupancy hardware conditions and overrides the static sequence length guards. By allowing a higher split count for low-head configurations, our approach increases Hopper SM occupancy. We observe kernel-level speedups of roughly 21 to 24\% on metadata-enabled target configurations, with no observed performance degradation in high-load scenarios.

\section{Background and Motivation}
\label{sec:background}

\subsection{Occupancy Collapse in Low-Head-Count Decoding}
Decode-step attention is inherently a memory-bound reduction over the sequence dimension. When $H_{KV} = 1$ (MQA), or more generally when $H_{KV}$ is small in GQA, the total number of attention work tiles ($Batch \times H_{KV}$ for decode, where $L_Q=1$) can be drastically smaller than the available SM count. Given $132$ SMs on an H100 GPU \cite{nvidia2022hopper}, operating on 8 tiles without sequence splitting translates to an occupancy of sequence computing blocks of approximately 6\%.

\subsection{The Premature Guard Flaw}
The standard FlashAttention-3 scheduling logic relies on a parameter $s$ (number of splits) to distribute the workload along the sequence dimension. However, an explicit guard in the underlying C++ heuristic returns $s=1$ if the sequence length $L_K \le 512$, assuming the splitting overhead outweighs the computational benefit. This static threshold overlooks the hardware scale of H100 and the low head count of MQA and low-head-count GQA, preventing the workload optimizer from considering higher split factors that would mitigate SM underutilization.

\section{Automated Discovery via Evolutionary Search}
\label{sec:discovery}

The limitations of the standard heuristic were initially exposed using \texttt{OpenEvolve} \cite{openevolve2024}, an LLM-guided evolutionary search framework. Rather than manually tuning C++ kernel parameters, we optimized the workload scheduling directly through the FlashAttention-3 Python interface. This approach allowed an iterative search agent to dynamically generate, evaluate, and refine Python-based heuristics in-the-loop on a live H100 GPU.

\subsection{Experimental Design and Search Space}

The experiment was designed to minimize Time per Output Token (TPOT) for standard chat interactions, targeting Llama-3.1-70B-Instruct \cite{ai2024llama3} with short prompts ($L_K \le 512$, $Batch=1$). On architectures like the H100, traditional heuristics are tuned for high-throughput heavy workloads, whereas short sequence decoding is bounded by kernel launch overhead and low occupancy.

We isolated the scheduling semantics from the mathematical correctness of attention by exposing three primary parameters to the evolutionary agent:
\begin{enumerate}
    \item \texttt{num\_splits} (Split-KV): Controls sequence-level parallelization across SMs.
    \item \texttt{pack\_gqa}: A boolean flag managing memory layouts for Grouped Query Attention (Llama 70B uses an 8:1 Key-Value ratio).
    \item \texttt{sm\_margin}: An integer governing resources reserved for the Cooperative Thread Array (CTA) scheduler for final reductions.
\end{enumerate}
Variables defining model semantics, such as input tensors, causality constraints, sliding window sizes, and RoPE embeddings, were frozen so that numerical correctness was easier to preserve while the search explored only scheduling behavior.

\subsection{Evolutionary Process}

The evolutionary agent synthesized dynamic Python logic rather than static constants. The evaluation framework compiled and cached target variants via a subprocess evaluator, rejecting invalid or numerically unstable candidates.

During isolated microbenchmarking, the standard C++ heuristic enforced \texttt{num\_splits = 1} due to the short sequence length guard ($L_K \le 512$). Over subsequent generations navigating the non-convex parameter space, the algorithm identified a correlation: increasing \texttt{num\_splits} in low-throughput regimes directly correlates to latency reductions.

\begin{figure}[htbp]
\centering
\begin{lstlisting}[language=Python, frame=single, numbers=left]
if batch_size == 1:
    local_num_splits = 12  # Optimal for <500 range (TARGET)
    local_pack_gqa = True
    local_sm_margin = 0
    if seqlen_k < 256:
        local_num_splits = 16  # Max splits for very short
\end{lstlisting}
\caption{A fragment of the high-performing evolved Python heuristic.}
\label{fig:evolved_logic}
\end{figure}

The generated logic (Figure \ref{fig:evolved_logic}) bypassed the underlying static guard by forcing \texttt{num\_splits = 12} or \texttt{16} for short-prompt single-batch requests. 

\subsection{Empirical Dissection and Motivation}

Analysis of the top-performing evolutionary candidates revealed the physical mechanism behind the acceleration. In short-context, low-tile decode configurations, the static short-sequence guard keeps $s=1$ even when the resulting grid underfills the H100. The strongest evolved Python candidates repeatedly overrode that behavior by forcing much larger split counts, typically $s=12$ or $16$ for short single-batch prompts, thereby increasing parallel work across SMs and recovering latency.

We treat these aggressive evolved settings as evidence of the mechanism, not as the final deployed policy. The paper therefore evaluates a narrower C++ rule focused on the clean $nblk=4$ boundary bucket, reported through the representative $L_K=512$ case; extending the benefit to lower $L_K$ values and learning more configuration-specific split counts is future work.

\section{Sequence-Aware Split Heuristic}
\label{sec:methodology}

Our solution distills the evolutionary observation into a conservative C++ policy in \texttt{heuristics.h}, where both sequence length blocks (\texttt{nblk}) and total available work tiles (\texttt{total\_mblocks}) are available. In the decode-like regime studied here, \texttt{total\_mblocks} corresponds to the aggregate tile count; with $L_Q=1$, this reduces to the earlier $Batch \times H_{KV}$ intuition because there is only one M-block per head. To keep the demonstration easy to interpret, the policy changes behavior only in the low-tile $nblk=4$ boundary bucket, which we report through the representative $L_K=512$ case:

\begin{itemize}
    \item \textbf{Guard 1 (Short Contexts Left Unchanged):} If $nblk \le 3$ (e.g., $L_K \le 384$), keep $s=1$ in this initial policy.
    \item \textbf{Guard 2 (Saturated Boundary Case):} If $nblk = 4$ but the hardware is adequately saturated (e.g., $H_{KV} \ge 4$ leading to sufficient tiles), keep $s=1$.
    \item \textbf{Low-Tile Boundary Case (Current Demonstration):} If $nblk = 4$ and the SMs are starved (i.e., only a few tiles are available, such as $Batch \times H_{KV} < 4$), use a small conservative split count ($s=3$ on the current stack). The reported measurements focus on the representative $L_K=512$ case.
\end{itemize}

This policy isolates the core effect in the cleanest regime while leaving all other cases on their existing path.

\subsection{\texorpdfstring{Scope: Why $L_K=512$ and Not Shorter?}{Scope: Why L\_K=512 and Not Shorter?}}
\label{sec:scope}
Section~\ref{sec:discovery} exposed a broader phenomenon than the final paper policy: once the premature shortcut is bypassed, short low-tile decode cases can benefit from additional splitting. The policy evaluated here is intentionally narrower. Boundary-sweep measurements show unchanged behavior at $L_K \in \{128, 256, 384\}$, a clear win at the representative $L_K=512$ point within the $nblk=4$ boundary bucket, and unchanged behavior again once the baseline efficiency loop already runs for longer contexts (e.g., $L_K \ge 640$).

This should therefore be read as a conservative proof of concept, not as a claim that lower-$L_K$ cases can never benefit from splitting. We restrict the rule to the cleanest boundary regime so the paper can demonstrate the central idea with a simple, stable policy. Extending the benefit to lower $L_K$ values and learning more configuration-specific \texttt{num\_splits} values is future work.

\subsection{The C++ Heuristic Patch}

The discovery motivated a direct modification to the underlying \texttt{heuristics.h} source file within the FlashAttention-3 Hopper stack. The evolutionary method operated on the FlashAttention-3 Python interface, whereas the evaluated policy is expressed at the C++ \texttt{heuristics.h} level using the variables below. The original guard strictly enforced $s=1$ when \texttt{num\_n\_blocks} $\le 4$ ($L_K \le 512$). The paper policy keeps short-context and saturated cases unchanged, and introduces a single low-tile override for the $nblk=4$ boundary bucket, which we evaluate at the representative $L_K=512$ point:

\begin{figure}[htbp]
\begin{lstlisting}[language=C++, frame=single, numbers=left]
// Guard 1: L_K <= 384 (nblk <= 3) - leave shorter contexts unchanged
if (num_n_blocks <= 3) { return 1; }

// Guard 2: nblk = 4 boundary bucket with enough tiles
// total_mblocks is the aggregate work-tile count; for decode (L_Q = 1),
// this reduces to batch_size * num_heads_kv.
if (num_n_blocks <= 4 && total_mblocks >= 4) { return 1; }

// Low-tile boundary case: demonstrate the idea with one small override
if (num_n_blocks == 4 && total_mblocks < 4) { return 3; }

// For longer contexts, existing efficiency loop runs (unchanged)
\end{lstlisting}
\caption{Conservative C++ policy used in the paper: keep shorter and saturated cases unchanged, and override the low-tile $nblk=4$ boundary bucket with $s=3$.}
\label{fig:cpp_patch}
\end{figure}

This policy leaves shorter sequences ($L_K \le 384$) and saturated workloads (where tiles $\ge 4$) untouched, and adds one explicit override in the low-tile $nblk=4$ regime. In the reported measurements, the representative $L_K=512$ case within that bucket uses $s=3$ for $Batch=1$ and $H_{KV}\in\{1,2\}$, which is enough to demonstrate the benefit of sequence-aware splitting without introducing a broader configuration-specific policy surface.

\textbf{Bridging the Python and C++ Split Gap:} OpenEvolve discovered that aggressive split counts can recover latency once the premature shortcut is bypassed. The paper distills that observation into a much simpler rule: preserve unchanged cases and add one small override ($s=3$ on the current stack) in the cleanest low-tile boundary regime. Lower-$L_K$ extensions and more configuration-specific split choices are future work.

\section{Empirical Evaluation}
\label{sec:evaluation}

We evaluated the patched C++ kernel against the unpatched upstream FlashAttention-3 binary. To isolate framework-level dispatch overheads (e.g., PyTorch overhead), we used CUDA Graph replay and A/B-interleaved timing within the Python bindings to measure pure kernel execution times.

\subsection{Kernel-Level Speedups (A/B Testing)}

We benchmarked identical workloads representing common autoregressive decoding shape targets, where a shape denotes the tuple $(Batch, L_Q, L_K, H_Q, H_{KV}, D)$. Llama 3 70B is a GQA model with $H_{Q}=64$ and $H_{KV}=8$ \cite{ai2024llama3}; under 8-way tensor parallelism this maps to $H_{KV}=1$ per device, placing the per-device decode kernel in the same low-head-count regime as MQA.

\textbf{Precomputed scheduler metadata.} The results in Table~\ref{tab:main_results} are measured with \emph{precomputed scheduler metadata} (\texttt{get\_scheduler\_metadata()}) and explicit \texttt{num\_splits} passed from the Python bindings. This is the path used by inference stacks (e.g., vLLM) that precompute scheduling metadata before kernel launch. In that deployment path, the benchmark passes the split selected by each policy explicitly at launch time, so the A/B comparison measures the metadata-enabled behavior that an upstreamed heuristic would induce. Without precomputed metadata, the kernel uses an internal heuristic path and yields more modest gains ($\sim$1.0 to 1.05$\times$). The full 21 to 24\% improvement therefore applies only to deployments that already use or adopt the scheduler metadata API. The results, summarized in Table~\ref{tab:main_results}, demonstrate scaling improvements.

\begin{table}[H]
\centering
\caption{Kernel Testing for $Batch=1$ across $H_{KV}\in\{1,2,8\}$: Standard vs. Sequence-Aware Patched Kernel (\texttt{BF16})}
\label{tab:main_results}
\begin{tabular}{llccc}
\toprule
$L_K$ (Sequence Length) & $H_{KV}$ (KV Heads) & Standard (\si{\us}) & Patched (\si{\us}) & Speedup \\
\midrule
128 & 1 & 9.56 & 9.56 & 1.00$\times$ \\
128 & 2 & 9.45 & 9.45 & 1.00$\times$ \\
128 & 8 & 9.46 & 9.46 & 1.00$\times$ \\
256 & 1 & 11.57 & 11.57 & 1.00$\times$ \\
256 & 2 & 11.58 & 11.58 & 1.00$\times$ \\
256 & 8 & 11.60 & 11.60 & 1.00$\times$ \\
384 & 1 & 13.60 & 13.60 & 1.00$\times$ \\
384 & 2 & 13.57 & 13.57 & 1.00$\times$ \\
384 & 8 & 13.55 & 13.55 & 1.00$\times$ \\
\textbf{512} & \textbf{1} & \textbf{13.72} & \textbf{11.37} & \textbf{1.21$\times$} \\
\textbf{512} & \textbf{2} & \textbf{13.52} & \textbf{10.93} & \textbf{1.24$\times$} \\
512 & 8 & 13.56 & 13.56 & 1.00$\times$ \\
2048 & 1 & 11.99 & 11.99 & 1.00$\times$ \\
2048 & 2 & 12.66 & 12.66 & 1.00$\times$ \\
2048 & 8 & 12.73 & 12.73 & 1.00$\times$ \\
4096 & 1 & 13.88 & 13.88 & 1.00$\times$ \\
4096 & 2 & 13.53 & 13.53 & 1.00$\times$ \\
4096 & 8 & 15.05 & 15.05 & 1.00$\times$ \\
\bottomrule
\end{tabular}
\end{table}

The evaluated policy uses a higher fractional split factor ($s=3$ on the current stack for the reported $L_K=512$ case with $H_{KV}\in\{1,2\}$), increasing the active CTA count and SM utilization. The $L_K=2048$ and $4096$ rows are included as unchanged controls: the new override affects only the $nblk=4$ boundary bucket, while longer contexts fall through to the pre-existing efficiency loop. This yields roughly 21 to 24\% execution-time reductions for the target low-head-count shapes.

\subsection{Extended Split Sweep and the Chosen Split Count}

\begin{figure}[htbp]
\centering
\includegraphics[width=0.72\textwidth]{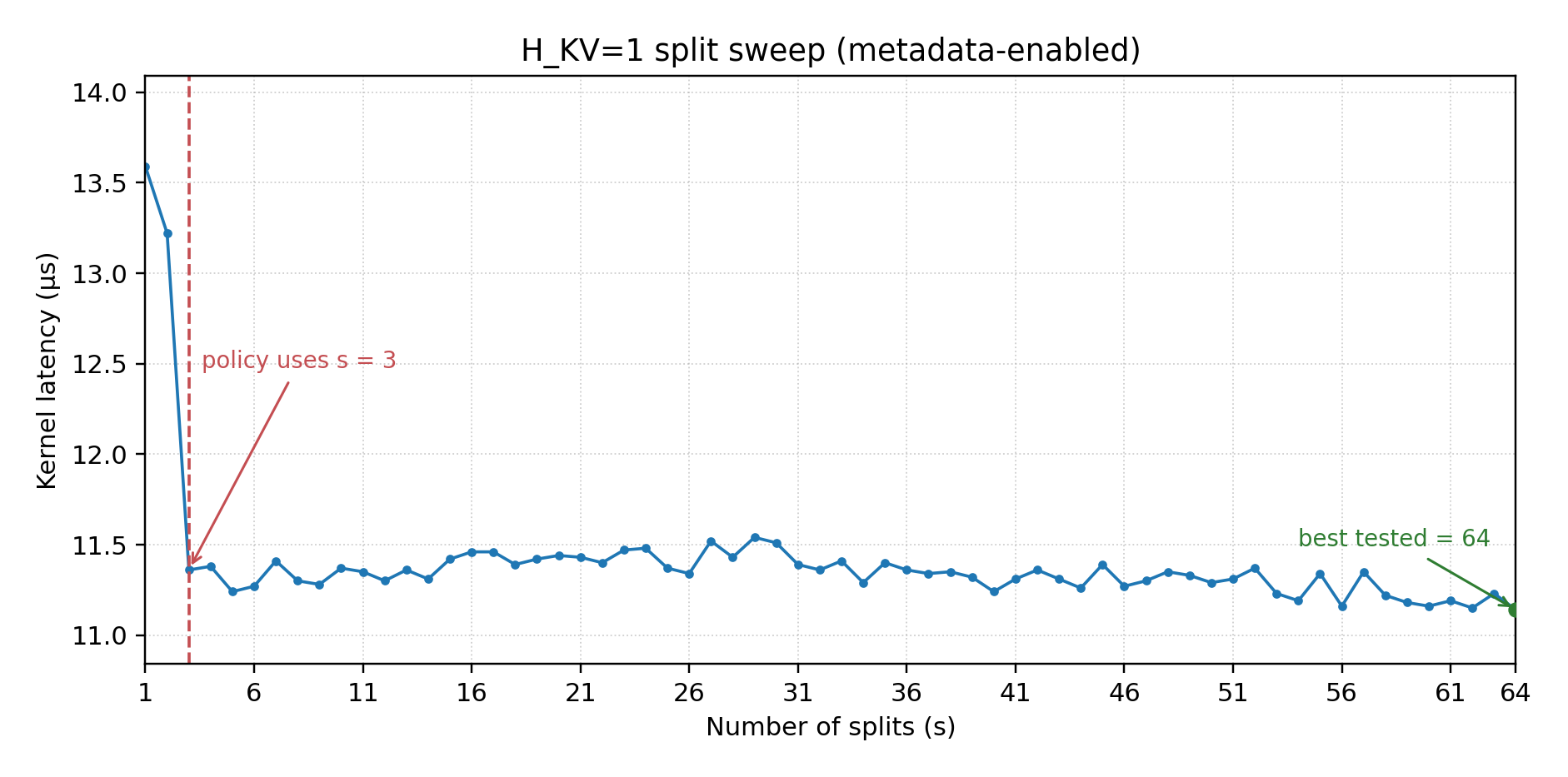}
\caption{Extended kernel-level split sweep for $Batch=1$, $L_K=512$, $H_{KV}=1$, and $D=128$ using precomputed scheduler metadata. Latency drops sharply once sequence splitting is enabled and then flattens into a broad low-latency plateau with shallow local minima at larger split counts. The best tested value in this sweep is $s=64$ ($\sim$11.14\,\si{\us}); the policy uses $s=3$ as a safeguard (smallest split entering this regime), with gain under $\sim$2\%.}
\label{fig:u_curve}
\end{figure}

To understand why the paper evaluates a small split count rather than an aggressive one, we extended the metadata-enabled split sweep for the boundary case ($L_K = 512$, $Batch=1$, $H_{KV}=1$) from $s=1$ to $s=64$. Figure \ref{fig:u_curve} plots the resulting kernel latencies. The sweep shows a steep improvement from $s=1$ into a broad low-latency plateau, with shallow local minima continuing beyond $s=8$.
\begin{itemize}\raggedright
    \item \textbf{Under-split ($s=1$):} Kernel execution is $13.72\si{\us}$ because only 8 CTAs are active, leaving most H100 SMs idle.
    \item \textbf{Low-latency plateau ($s \ge 3$):} Once splitting is enabled, execution time falls into a narrow band near 11.2 to 11.5~\si{\us}. For $H_{KV}=1$, the patched kernel achieves $11.37\si{\us}$ at $s=3$; for $H_{KV}=2$, $10.93\si{\us}$ at $s=3$.
    \item \textbf{Chosen split ($s=3$):} The paper uses $s=3$ as a safeguard: it is the smallest split that enters the low-latency regime. The extended sweep to $s=64$ (Figure~\ref{fig:u_curve}) shows the best tested value at $s=64$ ($\sim$11.14\,\si{\us}), but the gain from $s=3$ to the best is under $\sim$2\%. We keep one small override so the main effect is easy to attribute. Future work could extend the same idea to lower $L_K$ values and use more configuration-specific split choices.
\end{itemize}

\subsection{Safety and Regression Profiling}

To assess the risk of performance regressions, we conducted a sweep over 160 configurations spanning $Batch \in \{1, 2, 4, 8\}$, $L_K \in \{128, 256, 384, 512, 1024, 2048, 4096, 8192\}$, and $H_{KV} \in \{1, 2, 4, 8, 32\}$.

The empirical results showed no performance regressions across all configurations ($\ge 0.99\times$ standard). At $L_K=512$, wins appear only for $H_{KV}\in\{1,2\}$; the $H_{KV}\in\{4,8,32\}$ cases remain unchanged because both heuristics resolve to $s=1$. For dense configurations where splitting introduces atomic combination overhead (e.g., $Batch=8, H_{KV}=8$), the sequence-aware guard defaults back to $s=1$, matching standard execution time.

\section{Open Source Reproducibility}
\label{sec:reproducibility}

The reproduction package for this investigation, including the patch, benchmarking harnesses, and regression test matrix, has been open-sourced to aid verification.

The repository provides tools to compile standard and patched \texttt{flash\_attn\_3} kernels and run evaluations via Python bindings using CUDA Graph replays. Reviewers can reproduce the speedups on H100 hardware and verify the regression results by executing the test suite. The U-curve figure (Figure~\ref{fig:u_curve}) is generated from the \texttt{u\_curve\_sweep} experiment, which performs a kernel-level split sweep from $s=1$ to $s=64$ with precomputed scheduler metadata. Repository: \url{https://github.com/mllopartbsc/fa3-heuristic-fix}.

\section{Conclusion}
\label{sec:conclusion}

Adding a sequence-aware condition to the scheduling heuristic in FlashAttention-3 mitigates SM underutilization in low-head-count decoding. By considering both sequence length and tile count, inference latency for short sequences can be reduced while preserving standard performance in higher-throughput workloads. More broadly, this case study illustrates how OpenEvolve can uncover actionable systems-level optimizations that can then be distilled into small, upstreamable changes.

\bibliographystyle{unsrtnat}
\bibliography{refs}

\end{document}